\documentclass[pra,twocolumn,showpacs,superscriptaddress,groupedaddress,showpacs,shokeys]{revtex4} 
\usepackage{graphicx}
\usepackage{epstopdf}
\usepackage{amsmath,amsthm,amssymb}
\usepackage{color}
\usepackage{amsfonts}
\usepackage{bm}
\usepackage{url}
\usepackage{cleveref}
\usepackage{multirow}
\usepackage[caption=false]{subfig}
\draft 

\begin{document}


\title{A novel pulsed magnet for magnetic linear birefringence measurements} 



\author{J. B\'eard}
\affiliation{Laboratoire
National des Champs Magn\'etiques Intenses (UPR 3228,
CNRS-UPS-UGA-INSA), F-31400 Toulouse Cedex, France}
\author{J. Agil\footnote{jonathan.agil@lncmi.cnrs.fr}}
\affiliation{Laboratoire
National des Champs Magn\'etiques Intenses (UPR 3228,
CNRS-UPS-UGA-INSA), F-31400 Toulouse Cedex, France}
\author{R. Battesti}
\affiliation{Laboratoire
National des Champs Magn\'etiques Intenses (UPR 3228,
CNRS-UPS-UGA-INSA), F-31400 Toulouse Cedex, France}
\author{C. Rizzo}
\affiliation{Laboratoire
National des Champs Magn\'etiques Intenses (UPR 3228,
CNRS-UPS-UGA-INSA), F-31400 Toulouse Cedex, France}


\date{\today}

\begin{abstract}
In this paper,  we describe a novel pulsed magnet, called Foil Coil, which can deliver a field transverse to the light propagation of more than 10~T over about 0.8~meters operating without cryogenic equipment. It has been designed for linear magnetic birefringence measurements. We report on testing the coil, and also show some physics data taken in vacuum during its commissioning in the framework of the BMV apparatus, with special attention to noise induced by the pulse itself. Finally, we compare the preliminary results obtained here, with data from the previous BMV coil. \end{abstract}


\maketitle 

\section{Introduction}

Magnetic linear birefringence is a non linear optics effect that has been separately discovered in 1901 by Kerr \cite{Kerr} and in 1902 by Majorana \cite{Majorana}. This effect is also known as the Cotton-Mouton effect, since between 1905 and 1907, Cotton and Mouton have published at least 21 papers concerning this effect in liquids and colloidal solutions \cite{Cotton}. Due to the smallness of the effect, first measurements in gas have been performed only in 1933 \cite{Tsai1933} and 1934 \cite{Cotton1934}. Indeed, the first observation of the effect in a helium gas, the species showing the smallest effect, has been reported only in 1991 \cite{Cameron1991}.

A linearly polarized light, propagating in a region where a uniform magnetic field perpendicular to the light wave vector exists, becomes elliptically polarized. In the case of gases, where $\Psi \ll 1$ and the index of refraction $ n \approx 1$, the acquired ellipticity $\Psi$ can be written as
\begin{equation} \label{Psi}
\Psi = \pi \frac{L_B}{\lambda}\Delta n \sin(2\theta),
\end{equation}
where $L_B$ is the length of the region where the magnetic field is present, $\lambda$ is the light wavelength, $\Delta n = n_\parallel - n_\perp$ is the magnetic induced birefringence \emph{i.e.} the difference between the index of refraction for a light polarized parallel to the magnetic field, $n_\parallel$, and one polarized perpendicular to the magnetic field, $n_\perp$, $\theta$ is the angle between the light polarization and the magnetic field of amplitude $B$.

A characteristic of magnetic linear birefringence is that the effect is proportional to the square of the magnetic field amplitude.
\begin{equation}\label{KCM}
\Delta n = K_{CM}B^2
\end{equation}
where $K_{CM}$ is a constant depending on the gas species, pressure and temperature, and on the light wavelength.  $K_{CM}$  has been theoretically studied, and measured for a large number of gases \cite{RizzoRizzo1997}. For example, for one atmosphere of helium gas, at 0$^\circ$C, $K_{CM} = 2.4\times 10^{-16}$~T$^{-2}$ (see \emph{e.g.} \cite{Cadene2013} and references therein).

Non linear electrodynamics, a theoretical framework that contains classical electrodynamics as the lowest order, and Quantum ElectroDynamics (QED) as a special case, predicts that a linear magnetic birefringence should also appear when light propagates in a vacuum in the presence of a transverse magnetic field \cite{Fouche2016}. The QED prediction is that $K_{CM} = 4\times 10^{-24}$~T$^{-2}$ \cite{Battesti 2013} \emph{i.e.} that a vacuum induces the same effect as around $5\times 10^{17}$ helium atoms per m$^3$ at standard conditions.

Vacuum linear Magnetic Birefringence (VMB) measurements are therefore important tests of the standard model physics and also of physics beyond the standard model (see \emph{e.g.} \cite{HIMAFUN}). Vacuum $K_{CM}$ remains to be measured, notwithstanding more than a century of different attempts to observe it \cite{Battesti 2013}. To date, the best limit has been obtained with the PVLAS experiment \cite{PVLAS2020} using as the source of the magnetic field two permanent magnets, giving $K_{CM} = (1.9\pm8.1)\times 10^{-23}$~T$^{-2}$ at 99.7 $\%$ confidence level \emph{i.e.} a coverage factor $k=3$ \cite{CoverageFactor}\cite{Uncertainties}. This confidence level will be used throughout this paper. 

In 2001 a novel proposal has been put forward, \cite{BMV2001} and consequently an experiment, called \textit{Bir\'efringence Magn\'etique du Vide} (BMV), has been installed at the \textit{Laboratoire National des Champs Magn\'etiques Intenses} (LNCMI), Toulouse, France. More recently, another experiment, called Observing Vacuum with Laser (OVAL), based in Tokyo, Japan, also using a pulsed magnet, has been mounted \cite{OVAL}. As far as we know, BMV and OVAL are the only attempts to measure VMB which are operational at the moment. The latest BMV results have been published in 2014, with $K_{CM} = (8.3 \pm 8.0)\times 10^{-21}$~T$^{-2}$ \cite{Cadene2014}. Recently, OVAL has reported $K_{CM} = (0.3 \pm 1.6)\times 10^{-20}$~T$^{-2}$ \cite{OVALThesis}.

The difficulties of such a pulsed field approach to VMB can be summarized as follows. One needs to have a very sensitive interferometer, capable of measuring phase delays of a few  $10^{-9}$~radians, while injecting thousands of amperes in a coil driven by a power supply delivering hundreds of kJ's over a few milliseconds \emph{i.e.} a few $10^{8}$~W. Clearly, magnet field technologies, optics techniques, and how to couple them are the main subjects to be treated in the following. We start therefore by giving an overview of the different choices made over time as far the magnetic field production is concerned. We stress, that any improvement in linear magnetic birefringence measurements has been triggered by advances in the magnetic field technology. Then, we describe the novel pulsed magnet that we have developed, called Foil Coil, which can deliver a field transverse to the light propagation of more than 10~T over a distance of 0.8 meters, operating without cryogenic equipment. During the commissioning, we have also taken some physics data. We recall the main facts concerning our optical apparatus and our data analysis method, and we report on our investigation of the noise induced by the magnetic pulse itself. We also give results in term of VMB. The new results are indeed encouraging, and we explain why by comparing them with the previous BMV measurements \cite{Cadene2014}.

\section{The magnetic field}
If one considers equations \ref{Psi} and \ref{KCM}, the main experimental parameter to maximize the signal is the factor $B^2L_B$. In the case of a 1907 Cotton-Mouton measurement \cite{Cotton1907}, $B^2L_B$ was about 0.14~T$^{2}$m, corresponding to 1.85~T over $4.3\times 10^{-2}$~m, and the reported $K_{CM}\approx 10^{-7}$~T$^{-2}$.  To improve the measurements, improved magnets have always been needed. The first gas measurements of Tsa\"{\i} Belling \cite{Tsai1933} and Cotton and Tsa\"{\i} Belling \cite{Cotton1934} have been performed using the ``Bellevue'' electromagnet installation \cite{BellevueCNRS} in the Paris neighborhood. This instrument can be considered as one of the first physics facilities in the world. For the 1934 measurements, an extra coil was designed and used, increasing the maximum field from about 4.5~T to about 6.1~T over a distance of around 0.3~m, approaching 10~T$^{2}$m and reaching a $K_{CM}\approx 1.5\times 10^{-10}$~T$^{-2}$.

The theoretical aspects of this linear magnetic birefringence have been elucidate by Buckingham and Pople in 1956  \cite{BuckinghamTheory} and novel gas measurements have been performed in 1967 when Buckingham and collaborators renewed the field from the experimental point of view \cite{Buckingham1967}. The magnet used in the 1967 experiment was an electromagnet providing a field of about 2.8~T over half a meter \emph{i.e.} about 4 T$^{2}$m. The noise floor was about $K_{CM}\approx 1.5\times 10^{-15}$~T$^{-2}$ showing a considerable improvement with respect to the Cotton and Tsa\"{\i} Belling experiment even if the $B^2L_B$ of the two experiments was quite similar. This magnet was not so different from the one used in 1932 and 1940 by Farr and Banwell in one of the first attempts to detect any variation of the velocity of light in vacuum induced by the presence of a magnetic field \cite{Battesti 2013}.

In the 1980's, a new approach has been developed. Carusotto et al. introduced a modulation of the effect to be measured to increase the signal to noise ratio. This modulation was obtained by modulating the driving current of an electromagnet at about 0.4~Hz \cite{Carusotto1982} or by turning the entire magnet around its own axis at 0.9~Hz \emph{i.e.} changing the $\theta$ angle periodically (see formula \ref{Psi}) at 1.8~Hz frequency \cite{Carusotto1984}. The noise floor was about the same that the 1967 measurements \cite{Buckingham1967} but the $B^2L_B$ was less than 0.05~T$^{2}$m. Naively, to further decrease the noise floor one should have a $B^2L_B$ as high as possible, coupled with a frequency modulation as high as possible. This applies, in particular, to vacuum linear magnetic birefringence measurements which have been also put forward again in the 1970's \cite{Battesti 2013}. The design of VMB experiments are therefore more complicated than the gas one, and standard electromagnets are no longer the best choice for this kind of application.

Long superconducting dipole magnets have been developed for particle accelerators. In the 1990's, two of such magnets have been used by the BFRT (Brookhaven, Fermilab, Rochester, Trieste) collaboration \cite{Cameron1993} whose apparatus was installed at the Brookhaven National Laboratory at Upton, New York, USA. In the framework of this collaboration Cotton-Mouton effect of neon gas  \cite{Neon1991}, and helium gas \cite{Cameron1991} has been measured for the first time. The total magnet field length was 8.8~m, and the field was modulated at $3.2\times 10^{-2}$~Hz between 2.63~T and 3.87~T,  by modulating the driving current of the superconducting coils maintained at 4.7~K. The effective $B^2L_B$ parameter was about 35.5~T$^{2}$m. The $K_{CM}$ noise floor was about $1.5\times 10^{-20}$~T$^{-2}$ \cite{Cameron1993}. To reach such a low noise level, light was kept in the magnetic field region by a multipass optical cavity consisting of two mirrors on both sides of the magnets. Light passed up to about 550 times through the magnetic region before being analyzed.

The use of optical cavities increases the total path and the acquired ellipticity, but adds an extra level of complexity to this kind of experiment. Actually, the optics become challenging, for example, in the BFRT experiment light traveled about 5~km before exiting the magnetic field region.

The end of the BFRT attempt corresponds to the start of the PVLAS collaboration that, after 25 years of work, has achieved the best result to date \cite{PVLAS2020}. In the PVLAS experiment, the effect modulation has been obtained by rotating the entire magnet around its own axis as in ref.~\cite{Carusotto1984}. Initially a superconductive magnet providing 5~T over 1~m has been used. This was subsequently replaced by two permanent magnets, rotating at about 5~Hz, providing 2.5~T over twice 0.85~m \emph{i.e.} a $B^2L_B$ of about 10~T$^{2}$m \cite{PVLAS2020}. To increase the effect to be measured, a resonant optical cavity has been used. Light is therefore trapped in the magnetic field region for milliseconds on average. Millions of seconds of integration time have been necessary to reach a $K_{CM} = (1.9\pm 8.1)\times 10^{-23}$~T$^{-2}$ \cite{PVLAS2020}.

A novel approach based on the use of pulsed magnets has been suggested in 1998 \cite{Rizzo1997}. Pulsed magnets \cite{HIMAFUN} are electromagnets, typically cooled to liquid nitrogen temperature, that are driven by the discharge of a bank of capacitors. A field of up to 100~T has been obtained, \cite{HIMAFUN} over several milliseconds, in a bore of $\simeq 1$~cm diameter. The field is usually in the Faraday configuration, as in the case of a standard solenoid. The use of pulsed fields is interesting because they are very intense and rapidly modulated. In ref. \cite{Rizzo1997} the suggestion was to use a pulsed magnet having a radial access to have the field perpendicular to the light wave vector, as required to measure Cotton-Mouton effects. This kind of coil is called a split coil since it is obtained by splitting a pulsed solenoid coil in two, introducing a spacer between the two parts. Obviously, the field obtained at the level of the radial access is smaller than in the case when the solenoid is not split into two.

The most expensive part of a pulsed field installation is the power supply, essentially a bank of capacitors. Its size and therefore its cost depends on the energy to be injected into the pulsed coils. In the case of split coils, to create a transverse field for magnetic linear birefringence measurements, most of the energy delivered creates magnetic field in regions that light does not cross, while the effective $B^2L_B$ is somewhat limited by the bore dimensions of a few centimeters.

In the year 2000 proposal launching the BMV project \cite{BMV2001}, following directly the 1998 paper \cite{Rizzo1997}, a different design for the pulsed magnet was therefore proposed \cite{Askenazy{2001}}. We refer to it as a plate-coil since it is based on the idea to allow current flow through plates parallel to the region where the magnetic field is needed, thus the field obtained is perpendicular to the light wave vector by construction. In the framework of the BMV project, construction tests have been conducted at the LNCMI, Toulouse, France, where the BMV experiment is installed. In particular, in 2002 a field of about 2.5~T has been obtained using a plate-coil, \cite{Batut2002} however, these tests confirmed that the main technical difficulty was to transport the current from one plate to another, during the pulse, without destroying the connection itself.

Therefore, the BMV best results \cite{Cadene2014} to date have been realized using what we call an Xcoil \cite{Xcoil}. In this pulsed magnet design, to maintain the magnetic field in the region crossed by the light as far as possible, the two electromagnets with a racetrack shape are disposed in an X geometry. In the test phase, the version used in ref. \cite{Cadene2014} reached more than 14~T over a length of 0.13~m corresponding to 25~T$^{2}$m. The total duration of a pulse is a few milliseconds. The magnetic field reaches its maximum value within 2 ms which is of the same order of magnitude as the storage time of photons in the optical Fabry-Perot cavity that is used to increase the optical path in the magnetic field region. The optical cavity acts as a low pass filter and the ellipticity induced by the magnetic field is therefore not proportional to the square of the field but to the filtered square of the field \cite{Agil2021}. When the cavity filtering is taken into account, as shown in ref. \cite{Agil2021}, $B^2L_B$ is no longer the parameter to be optimized, because increasing $L_B$ involves increasing the cavity length, reducing its bandwidth, which in turn increases the filtering effect on $B^2$.

An even larger version of the Xcoil, called the XXL-coil, has been designed and constructed at the LNCMI, Toulouse, France \cite{XXLcoil}. This pulsed coil has reached a field higher than 30~T when a current higher than 27000~A is injected into it. This corresponds to more than 300~T$^{2}$m \cite{HIMAFUN}. This coil requires a large liquid nitrogen cryostat, and the overall dimensions are hardly compatible with the BMV table top experiment. The construction of such a coil has proved to be difficult to manage even in a framework of a European standard high magnetic field laboratory such as the French LNCMI.

For all these reasons, we decide to design and manufacture a novel pulsed magnet which could provide a high field without the need to be cooled. We called it Foil Coil. It is constructed using a single copper foil as explained in the following of this paper.

As already explained, the OVAL experiment is also based on a pulsed magnet. The latest version \cite{OVALThesis} consists of two race-track coils \cite{OVAL} in a Helmholtz configuration, cooled to liquid nitrogen temperature. This magnet operates at a peak value of field of 8.3~T over a field length of 0.17~m, and at a very high repetition rate of 0.05~Hz. The results reported in \cite{OVALThesis} have been obtained accumulating $2.6\times 10^4$ pulses.

\section{The Foil Coil}

\subsection{Design}

This new magnet has been designed with several constraints, some are imposed by physical or technical limitations, while others can be chosen to facilitate the experiment. Physical limitations are the same as the one encountered by standard pulsed magnets, namely the mechanical stress due to Lorentz forces and the heating due to Joule effect. The magnet optical access diameter is chosen to be 17~mm. The objectives in terms of magnetic field and pulse duration, are respectively in the range 10--20~T and 5--10~ms. This allow us to use copper as the conductor with an extra reinforcement outside the coil and to operate the magnet at room temperature. This is the major difference with other pulsed magnets, which are usually cooled in liquid nitrogen temperature. Avoiding the need for a nitrogen cryostat allows us to use the full optical access diameter of the magnet as there is no thermal insulation between the coil and the vacuum tube of the experiment.

As we stated previously, the most expensive and the least flexible part of a pulsed magnet system is the capacitor bank, thus the parameters of the magnet, such as the inductance or conductor section, have to be adapted to existing generators.  The capacitor bank that energized the Foil Coil is one of the 3~MJ mobile banks of LNCMI, with a capacity of 10~mF that can be charged up to 24~kV and can deliver 75~kA. The minimum rise time is 5~ms in short circuit so the rise time of the magnetic field must be higher to transfer some energy to the coil. Table~\ref{tab:paracoil} summarizes the properties of the magnet that result from all the constraints cited above.

\begin{table}
\centering
\begin{tabular}{|c|c|}
 \hline
Optical access (mm) & 17\\
  \hline
 Effective magnetic length (mm) & 820\\
  \hline
  Copper cross-section (mm$^2$) &\multirow{2}{*}{$72\times 0.5$}\\
    (height $\times$ thickness) &\\
  \hline
 Thickness of insulation ($\mu$m) & 120\\
  \hline
 Number of layers & 50\\
  \hline
 Efficiency (T/kA) & 0.537\\
  \hline
 Inductance ($\mu$H) & 700\\
  \hline
 Resistance at 26$^\circ$C  (m$\Omega$) & 48.1\\
  \hline
 Rise time (ms) & 5.8\\
\hline
\end{tabular}
\caption{Summary of the properties of the magnet installed on the BMV experiment.}
\label{tab:paracoil}
\end{table}

\subsection{Realization}

\begin{figure}
\includegraphics[width=8cm]{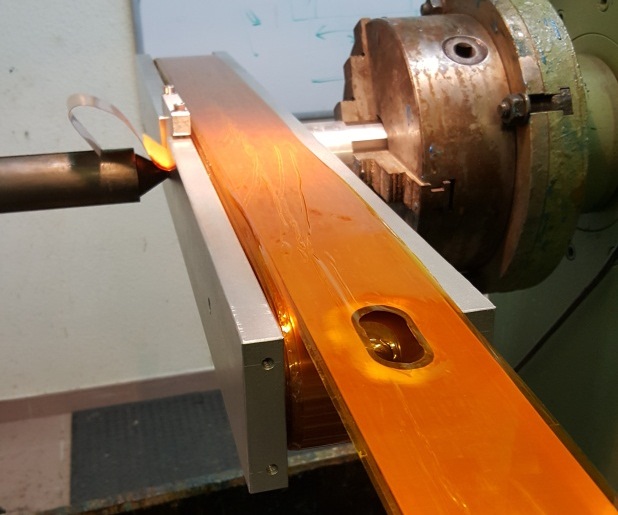}
\caption{Photo of the Foil Coil during the winding showing one hole in the copper foil and in the Kapton tape. The holes have a racetrack shape to keep a circular access when the foil is wound over the round end of the magnet. The copper is cut first, and then the insulation with Kapton tape is cut with a smaller hole to ensure a perfect electrical insulation between two layers thanks to the overlapping of the Kapton on the copper.}
\label{fig:FoilCoilWinding}
\end{figure}

The magnet is based on the winding of a copper foil insulated with two layers of Kapton tape.  About 100 meters of the copper foil is wound  over a glass fiber epoxy FR4 type body with a racetrack shape, representing 50 layers of conductor. The optical access is provided by two holes in each turn in the insulated copper. This design offers a very good symmetry of the generated magnetic field thanks to the homogeneous current distribution and the very small effect of the layer transition compared to a wire wound coil. Another difference with standard wire wound coils is the continuity of conducting metal perpendicularly to the axis of the laser beam that screens the axial component of the magnetic field avoiding the use of an extra shield. A image of the hole in the conductor and the insulation is shown in Fig.~\ref{fig:FoilCoilWinding}. The presence of the hole increase the current density by a factor 1.4 so the local temperature rise during the pulse is twice higher than in the rest of the coil.

\begin{figure}
\includegraphics[width=8cm]{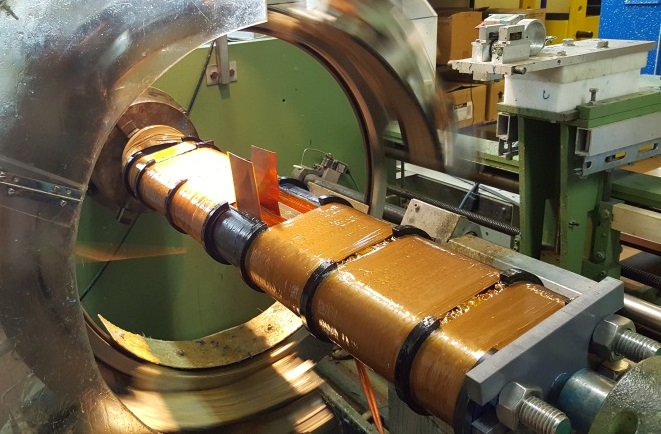}
\caption{Application of the mechanical reinforcement on the Foil Coil.  Two 3D printed halves of a hollow cylinder, in black on the photo, containing four M24 stainless steel tie rods and filled with epoxy resin permit to wind the Zylon fibers and to transmit the force in the direction parallel to the fibers. Zylon fibers are wound over a length of 600~mm with a thickness about 5~mm and are capable to sustain a force above 3000~tons giving us a safety margin of about a factor 4, but also limiting the possible displacement of the layers to around a few tens of micrometers, decreasing the risk of buckling of the foils. At each end of the magnet the mechanical stress is naturally transmitted as a hoop stress in the copper foil as the winding can be considered as two halves of a solenoid. The extremities of the magnet are reinforced with two 25 mm thick stainless steel plates that are also used to fix the magnet, screwed with the M24 tie rods.}
\label{fig:FoilCoilReinforcement}
\end{figure}

\begin{figure}
\subfloat[]{\includegraphics[height=4.5cm]{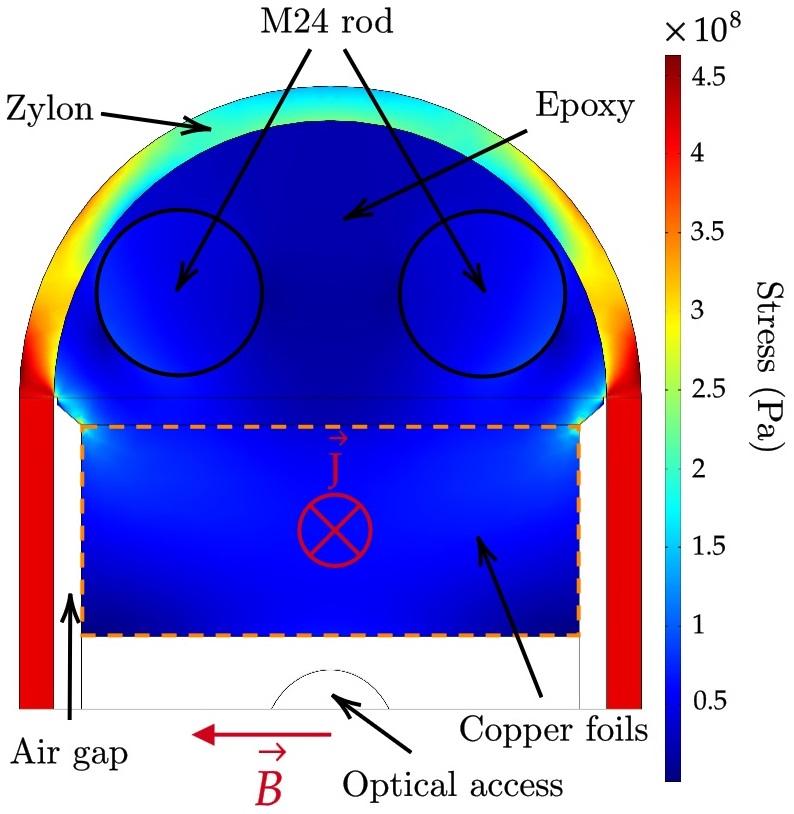}\label{fig:3a}}
\subfloat[]{\includegraphics[height=4.5cm]{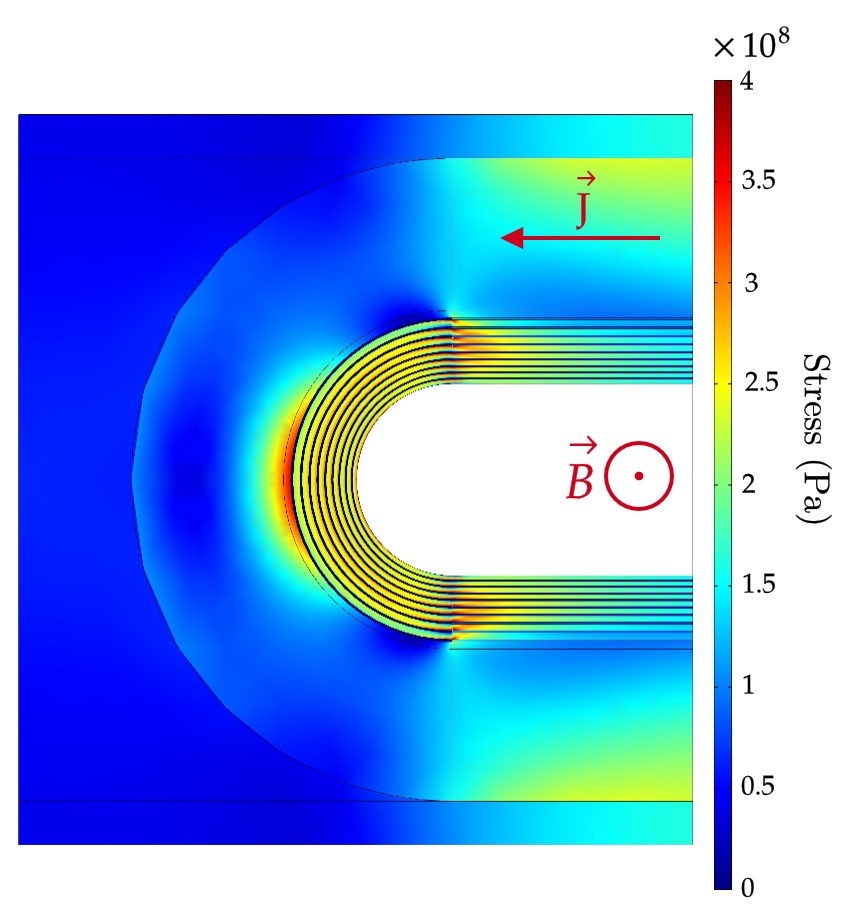}\label{fig:3b}}
\caption{FEA model of the mechanical stress in the Foil Coil during a 12.5~T pulse. (a) Cut along a plane parallel to $\protect\overrightarrow{B}$, $xy$-plane of Fig.~\ref{fig:FoilCoilFinish}, only the upper half is represented (b) Cut along a plane parallel to $\protect\overrightarrow{J}$ at one end of the magnet, $xz$-plane of Fig.~\ref{fig:FoilCoilFinish}. To save computing power only the first innermost layers are individualized taking into account their relative displacement. Indeed the first layers are self-supporting due to the increasing hoop stress with the radius of the layers up to half of the outer diameter. The outermost layers however behave as a whole as each layer is supported by the following one.}
\label{fig:FoilCoilComsol}
\end{figure}

The main limitation for the generation of high magnetic field is the huge stress due to Lorentz forces. In this particular design of a magnetic dipole, one component of the mechanical stress is due to the repulsion between each side of the winding. This force can be evaluated to several hundreds of tons and must be sustained by the reinforcement surrounding the coil. Most high field magnets are reinforced using Zylon fibers \cite{Toyobo}\cite{Herlach1991}, combining very high strength, low density and electrical insulation. We have decided to use this material, rather than metal, as the main reinforcement of the Foil Coil as shown in Fig.~\ref{fig:FoilCoilReinforcement}. Fig.~\ref{fig:3a} shows a Finite Element Analysis (FEA) model of the mechanical stress in a cross section of the Foil Coil in the plane parallel to the magnetic field. It shows the transmission of the forces from the foils to the Zylon fibres. At the ends of the magnet the mechanical stress is naturally transmitted as a hoop stress in the copper foils as another FEA simulation in Fig.~\ref{fig:3b} shows. Thanks to this solution containing the repulsive forces is no longer a problem. Furthermore, Zylon fibers are applied with a high pre-stress around 700~MPa that press the foils on the coil body preventing any risk of buckling in the straight parts of the winding. The limitation in terms of magnetic field comes from the potential buckling of the copper foils in the bent parts of the winding leading to the collapsing of the holes where it is not possible to place any efficient reinforcement. 

\begin{figure}
\includegraphics[width=8cm]{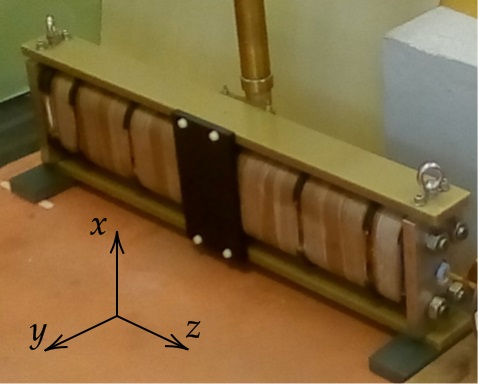}
\caption{Photo of the Foil Coil ready for tests. A reference frame is shown. The field is produced along the $y$ direction, Fig.~\ref{fig:3a} is cut along a $xy$-plane and Fig.~\ref{fig:3b} along a $xz$-plane.}
\label{fig:FoilCoilFinish}
\end{figure}

The magnet ready to be tested is shown in Fig.~\ref{fig:FoilCoilFinish}.

\subsection{Tests}
\begin{figure}
\includegraphics[width=8cm]{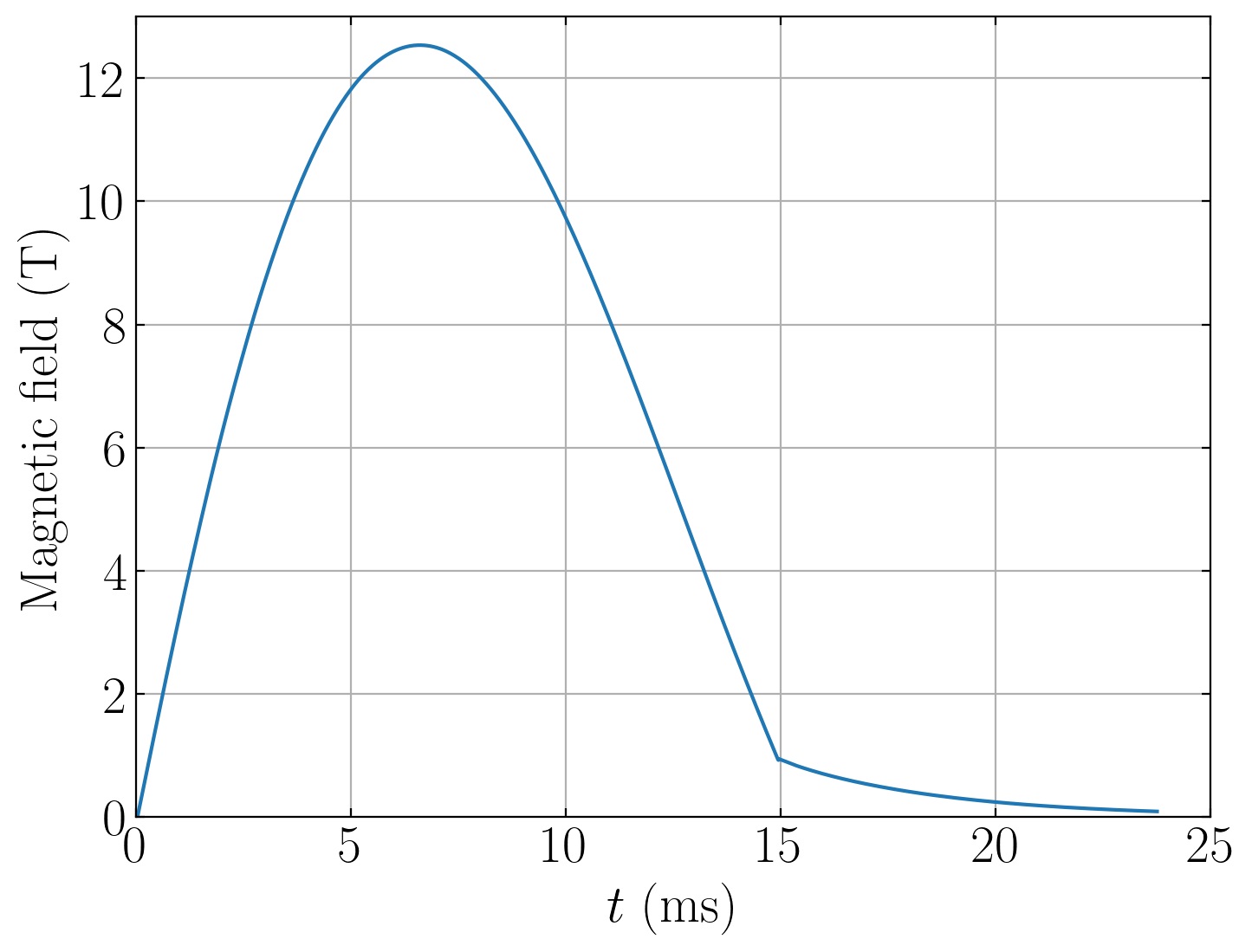}
\caption{Magnetic field temporal profile during a pulse at 12.5~T.}
\label{fig:FoilCoilShot}
\end{figure}

\begin{figure}
\includegraphics[width=8cm]{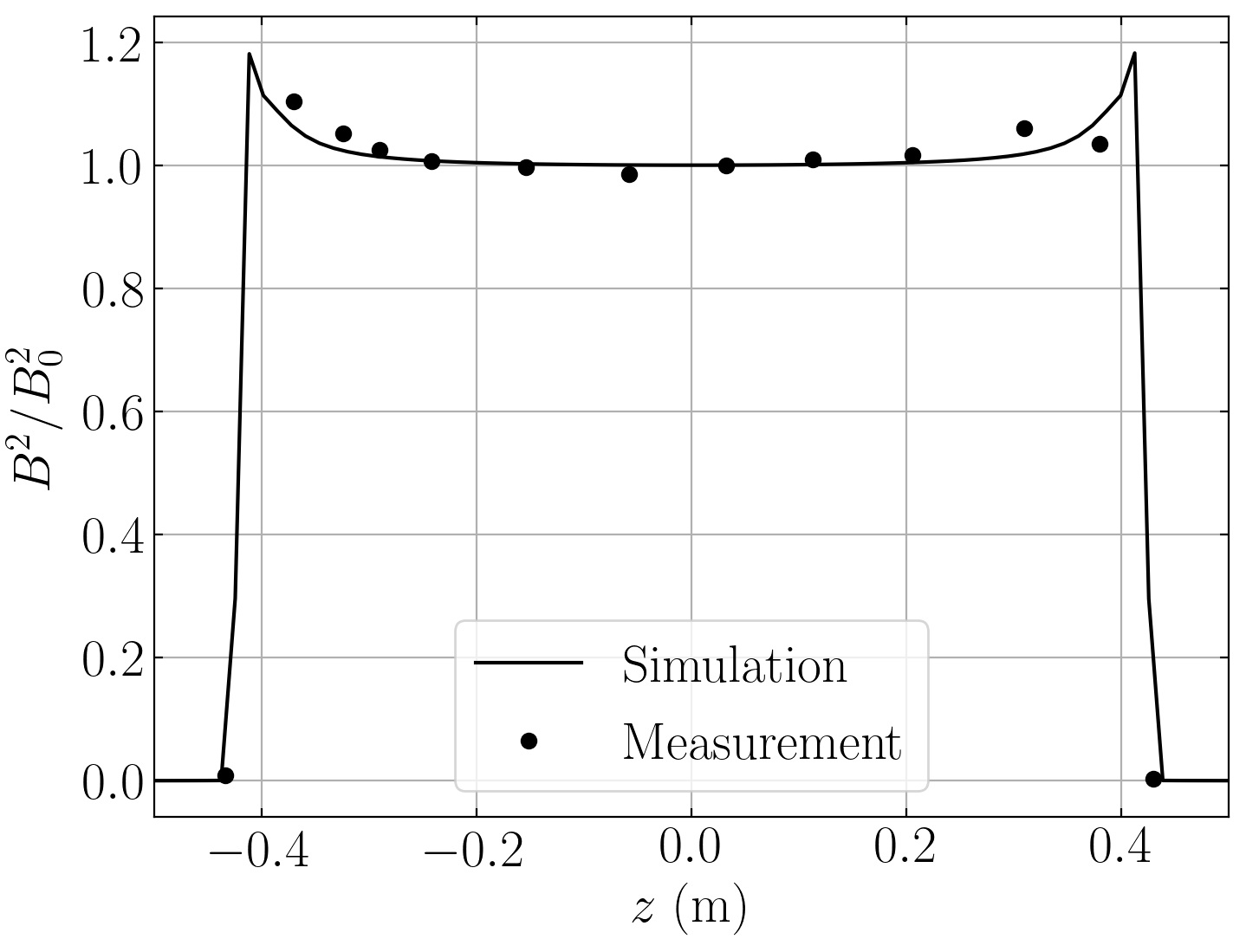}
\caption{Normalized profile of the magnetic field along the light axis produced by the Foil Coil. Solid black line present a finite element COMSOL simulation of the coil, black points show measurements.}
\label{fig:FoilCoilBprofile}
\end{figure}

Tests have been made up to 12.5~T, the obtained pulse profile is shown in Fig.~\ref{fig:FoilCoilShot} which corresponds to a $B^2L_B$ factor of 134~T$^2$m. This factor comes from integrating the magnetic field profile along the light axis as presented in Fig.~\ref{fig:FoilCoilBprofile}. During physics measurements, the maximum field is fixed to 11~T providing a reasonable safety margin. The effective maximum field achievable by the Foil Coil is hard to predict because buckling is difficult to simulate and is strongly influenced by small defaults in the winding. A next step will be to test this limit with a second prototype. However, if the Foil Coil can reach 20~T, this will lead to a temperature rise above 35~K per pulse, and cooling down will take more than two hours. Therefore, improvements in the cooling are required before testing the maximum field achievable. 

\begin{figure}
\includegraphics[width=8cm]{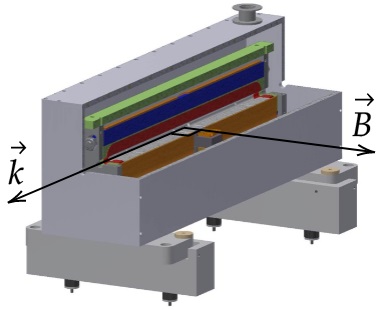}
\caption{CAD view showing a three quarter cut of the mechanical assembly. The vectors $\protect\overrightarrow{k}$ and $\protect\overrightarrow{B}$ respectively show the light propagation direction and the magnetic field direction, $\protect\overrightarrow{k}$ and $\protect\overrightarrow{B}$ are orthogonal. The foam inside the box has been removed. In red the winding generates a horizontal magnetic field. In blue the cylinder filled with epoxy resin, containing tie rods that fix the extremities of the coil. Around the blue cylinder, the 5~mm thick Zylon layer is visible.}
\label{fig:FoilCoilCAD}
\end{figure}

A pulsed magnet is submitted to huge internal forces during the pulse, and small movements or vibrations cannot be avoided. This vibrational noise will be transmitted to the experiment via the air and the mechanical support of the coil and the optical table. In order to reduce the noise generated by the vibration of the coil we decided to place it in a plastic box, with an interior covered by acoustic foam, as shown in Fig.~\ref{fig:FoilCoilCAD} . 

\begin{figure}
\includegraphics[width=8cm]{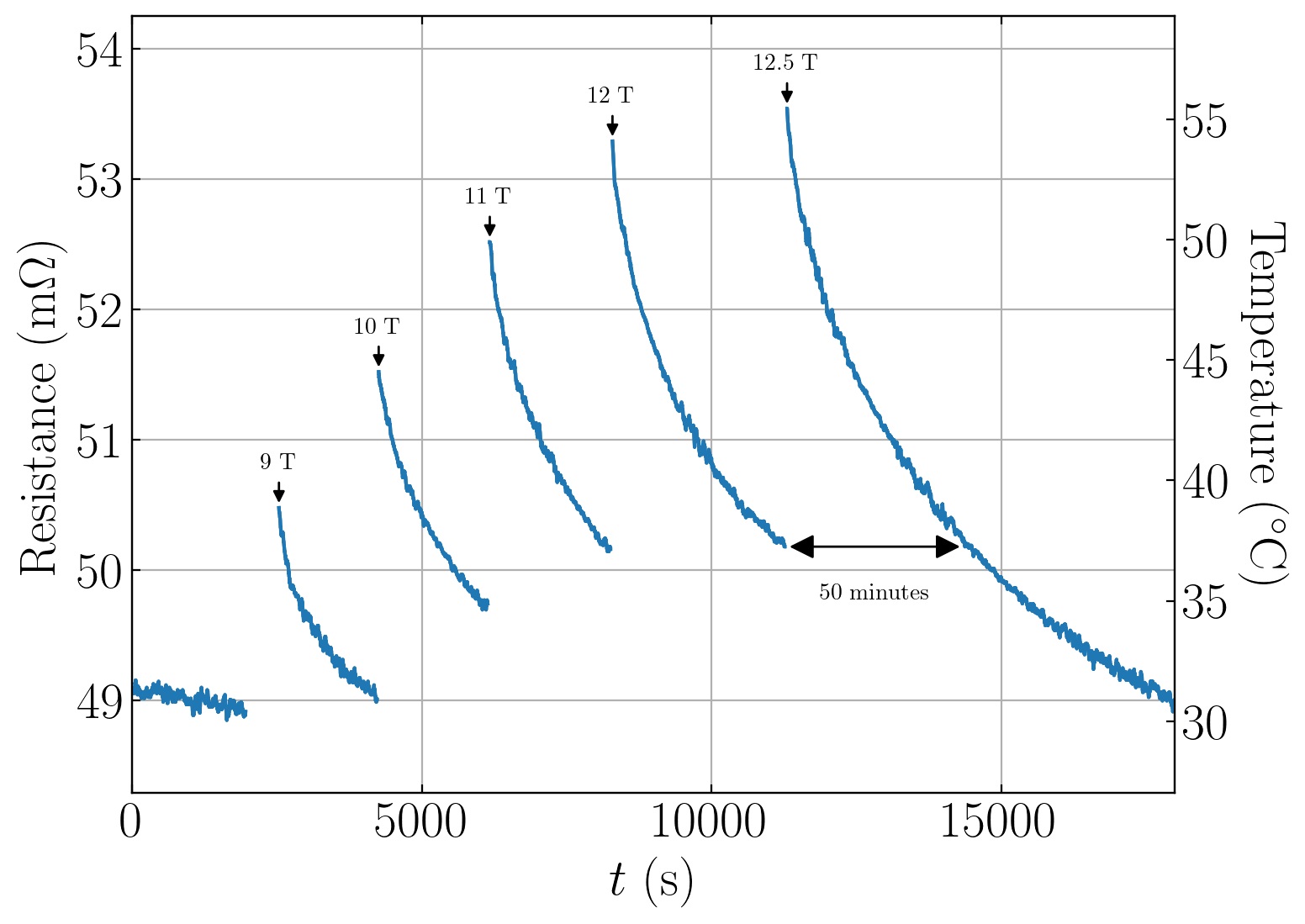}
\caption{Resistance of the Foil Coil measured during the tests up to the maximum field. Room temperature resistance is 48.1~m$\Omega$ and the maximum resistance is fixed at 54~m$\Omega$ representing a final temperature of the conductor about 60$^\circ$C.}
\label{fig:FoilCoilCooling}
\end{figure}

Repetition rate will be important to perform an experiment where hundreds of pulses will be necessary. This first prototype is far from having an optimized cooling because the main objective was to generate at least 100~T$^2$m at room temperature and in air. Fig.~\ref{fig:FoilCoilCooling} shows the evolution of the Foil Coil resistance, and therefore its temperature, after few pulses from 9 to 12.5~T during the magnet tests. The temperature of the Foil Coil can be deduced knowing the temperature dependence of Copper resistivity, which for our temperature range is linear $\rho=0.0668\times T+15.648$ with $\rho$ in n$\Omega$.m and $T$ in $^\circ$C, as confirmed by our measurements. Then measuring the Foil Coil resistance at any temperature, for example $R=48.1$~m$\Omega$ at 26$^\circ$C, we can deduced it for any other temperature given $R=\rho L/S$ with the $L/S$ factor calculated as $L/S=R(T=26^\circ\mathrm{C})/\rho(T=26^\circ\mathrm{C})$. Maximum average temperature of the conductor after a pulse is fixed to 60$^\circ$C corresponding to 54~m$\Omega$. Knowing that the temperature rises of about 22$^\circ$C, corresponding to a resistance increase of about 4~m$\Omega$ as shown in Fig.~\ref{fig:FoilCoilCooling}, after a pulse at maximum field a local temperature slightly above 80$^\circ$C is reached around the holes. This temperature is evaluated by knowing the cross section of conductor with and without the hole and considering that there are no skin effects or non-uniform current distribution. The presence of the hole decreases the section of conductor by about a factor of 1.4 leading to a temperature rise of about a factor of 2. A local temperature of 80$^\circ$C seems reasonable to preserve the insulation integrity and the mechanical properties of the whole assembly after thousands of pulses. 

\begin{figure}
\includegraphics[width=8cm]{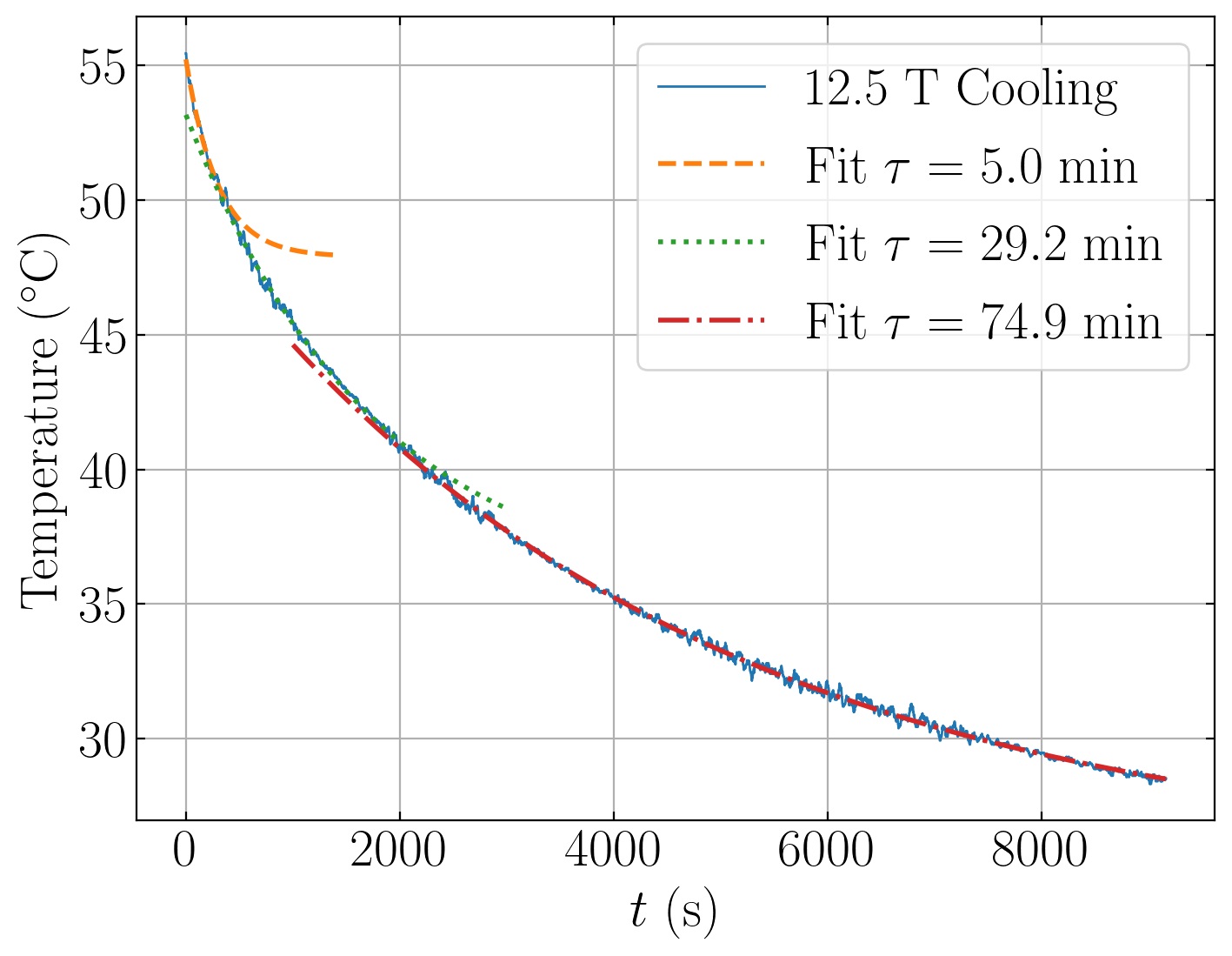}
\caption{Temperature of the Foil Coil of a function of time after a pulse at 12.5~T.}
\label{fig:FoilCoilShotCool}
\end{figure}

The time evolution of the cooling of the Foil Coil follows consecutive stages. Fig.~\ref{fig:FoilCoilShotCool} shows the cooling after the 12.5~T pulse, and alongside the data we present three exponential fits to the three stages. Doing so on the other pulses of Fig.~\ref{fig:FoilCoilCooling} leads to similar results hinting at a common cooling law. First, after the adiabatic heating of the copper during the pulse, the temperature is first homogenized along the copper foil and the heat is transmitted to the insulation layers. Next, the temperature of the highly insulating mechanical reinforcement will increase, and finally natural convection outside the box will slowly contribute to the cooling of the magnet. Cooling power will increase with the final temperature reached by the magnet, so the highest repetition rate corresponds to pulses to the maximum temperature. A final temperature fixed at 60$^\circ$C leads to a cooling duration of the magnet before another pulse at 12.5~T of about 50 minutes. Duty cycle will increase proportionally to the inverse of the square of the magnetic field.

\section{Commissioning physics measurements}

Once the Foil Coil has been tested, we have proceeded to its commissioning. We have installed the magnet in the apparatus which has been detailed in \cite{Hartman2019}. Following our previous work \cite{Hartman2019}, \cite{Agil2021}, we have first focused our attention to the noise induced by the pulse itself. We have investigated the impact on the optics of different pulse rise times, ranging from 2.9~ms to 5.8~ms, and different field amplitudes \emph{i.e.} the energy injected into the system.

\begin{figure}
\includegraphics[width=8cm]{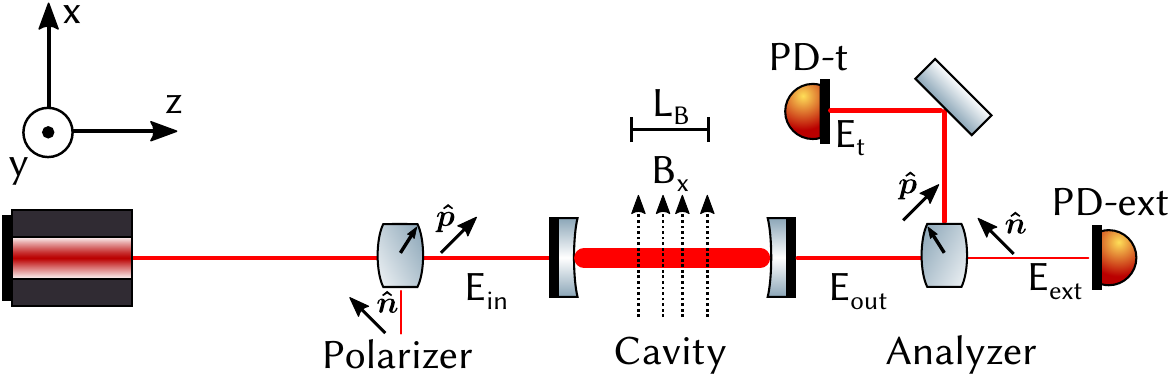}
\caption{Illustrative example of a vacuum magnetic birefringence experiment. The laser light is polarized before entry into the cavity surrounding the magnetic field region, after the cavity light is analyzed by another polarizer at extinction. Two photodiodes PD-t and PD-ext monitor respectively the ordinary beam $P_t$ and the extraordinary one $P_{ext}$.}
\label{fig-VMB_Polarimeter_cavity_enhanced}
\end{figure}

The principle of a polarimetric vacuum birefringence search is illustrated in Fig.~\ref{fig-VMB_Polarimeter_cavity_enhanced}.  In these setups, a laser field is linearly polarized in the $\textbf{p}=\cos\theta \textbf{x}+\sin\theta \textbf{y}$ direction. The angle $\theta$ is defined with respect to the direction of an applied transverse external field, $B(t)\textbf{x}$.

Due to the presence of a Fabry-Perot cavity of finesse $F$, as it propagates through the region of birefringent vacuum, the polarization of the light exiting the cavity acquires a total ellipticity $\Psi$ that is $2F/\pi$ larger than the one given in Eq.~\ref{Psi}
\begin{equation}\label{Psi2}
    \Psi = \frac{2F}{\lambda}K_\mathrm{CM}B^2 L_\mathrm{B}\sin(2\theta).
\end{equation}

The polarization components are then separated by the analyzer (power extinction ratio $\sigma^2$) at the output of the optical cavity.  After, the signal is recorded as the power at extinction of the second polarizer and compared to the total power exiting through the cavity end mirror,
\begin{equation}
	P_{ext} = \left[\sigma^2 + \left(\Gamma + \Psi\right)^2\right] P_t,\label{eqn:P_ext_polarimeter}
\end{equation}
where, taking into account that Fabry-Perot mirrors are also birefringent \cite{MirrorBir}, one has to introduce a corresponding ellipticity generated by the cavity itself, $\Gamma$. This ellipticity depends on various parameters in particular birefringence mirror axis orientation with respect to light polarization, light incident angle on each mirror surface, and intrinsic individual mirror birefringence \cite{MirrorBir}.

\begin{figure}
\centering
\includegraphics[width=8cm]{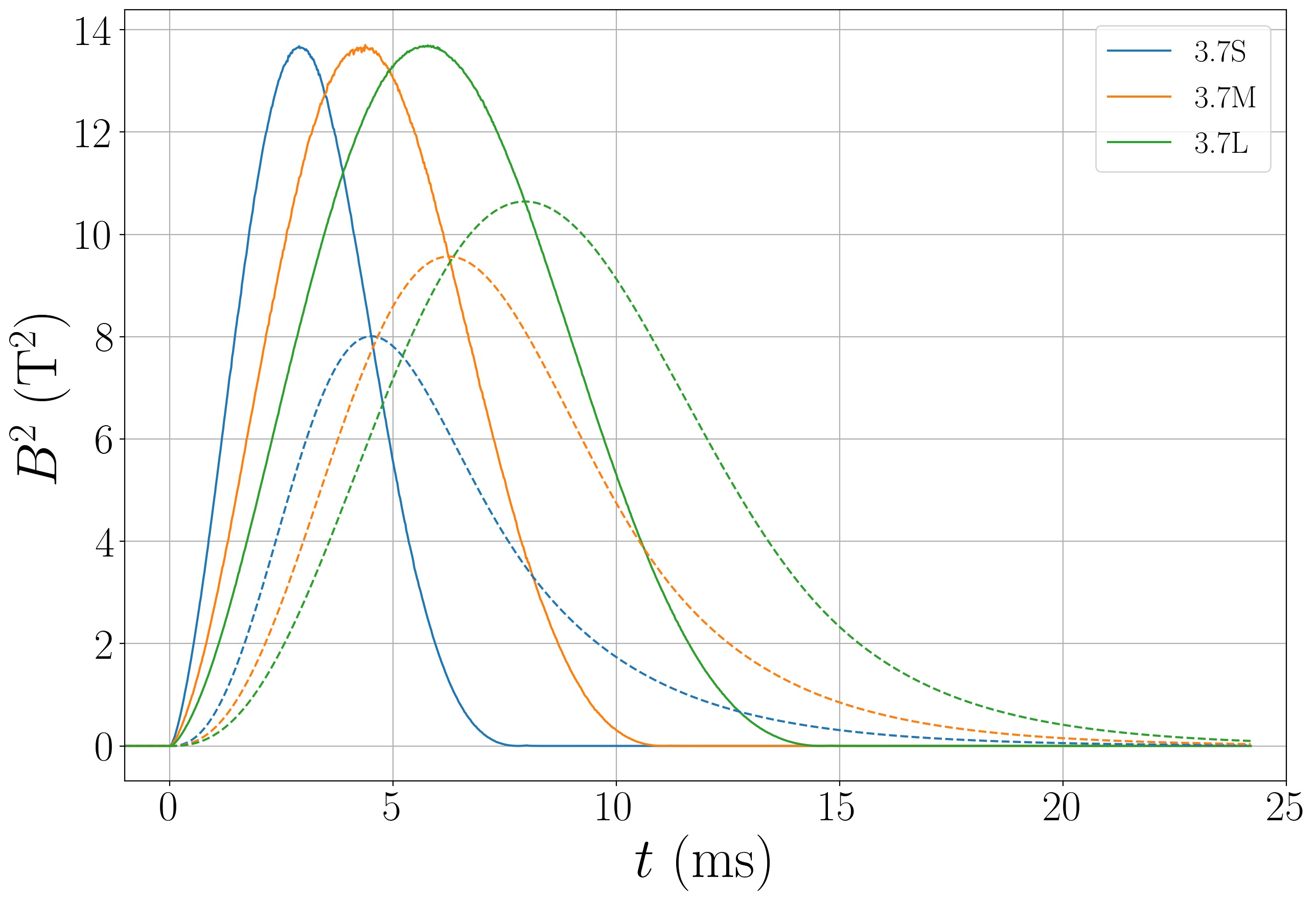}
\caption{Evolution of the square of the magnetic field in function of time for the three pulse lengths at 3.7~T. Solid line unfiltered pulse, dashed line pulse filtered by a low pass filter with a 55~Hz bandwidth corresponding to an optical cavity of 2.55~m length and a finesse of 535~000.}
\label{fig:FiltrChampsB2}
\end{figure}

\begin{figure}
\centering
\includegraphics[width=8cm]{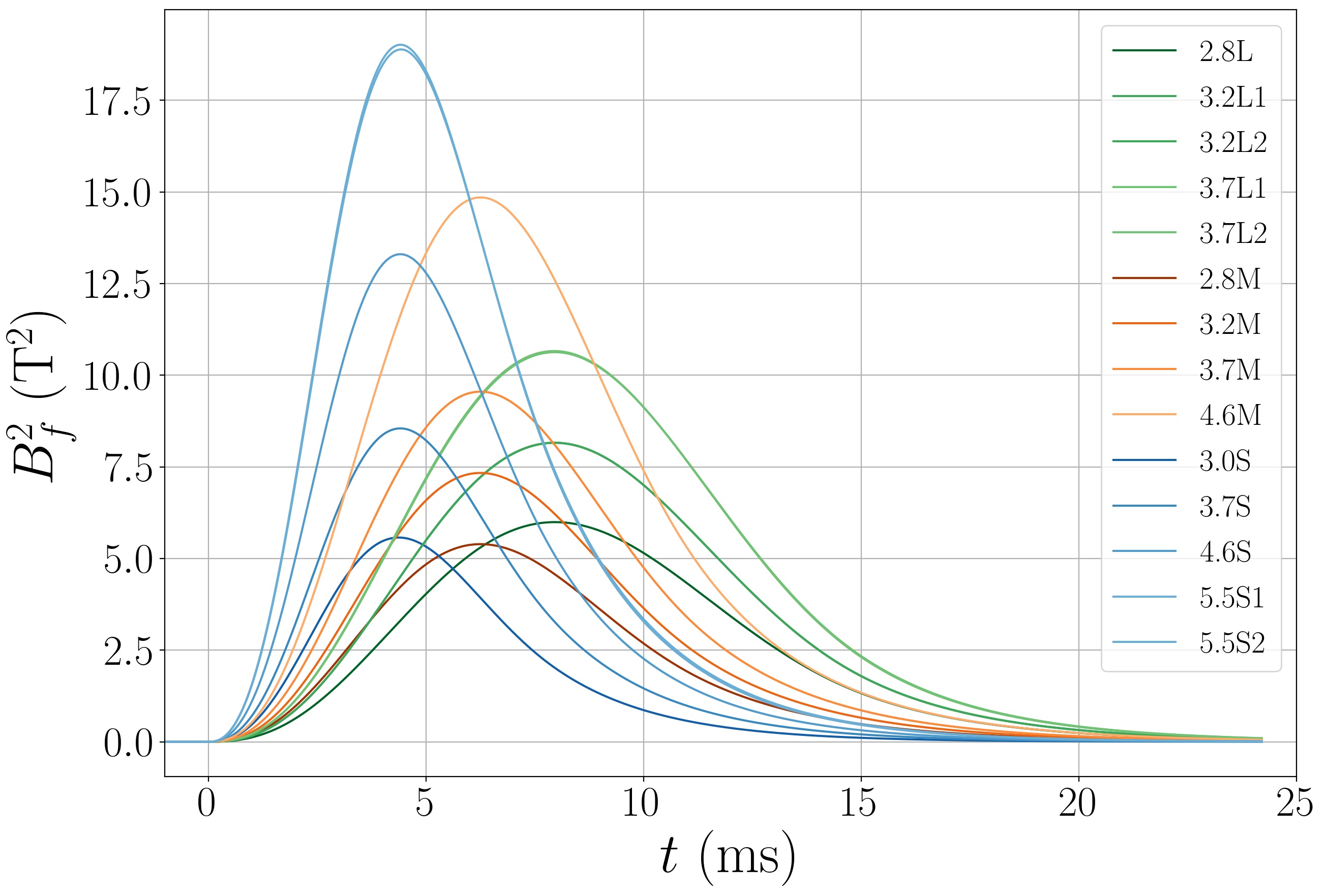}
\caption{Filtered square of the magnetic field as a function of time for every set.}
\label{fig:ChampsB2}
\end{figure}

The results reported here have been obtained from 578 pulses distributed over 14 sets with different pulse lengths and different pulse strengths. The different pulse lengths were achieved by disconnecting capacitors from the capacitor bank diminishing the period $T_0=2\pi\sqrt{LC}$ of the circuit comprising the coil and the capacitors. The first sets of pulses were with the full bank with a rise time of 5.8~ms. The second sets of pulses were with roughly half the capacitors so with a rise time diminished by a factor near $\sqrt{2}$ ending with a rise time of 4.4~ms. The final sets were with a quarter of the capacitors with a rise time halved to a value of 2.9~ms. To exemplify the effect of the cavity filtering as a function of the rise time of the magnetic field, we show in Fig.~\ref{fig:FiltrChampsB2} the filtered and unfiltered square of the magnetic field for the three pulse lengths at 3.7~T. In Fig.~\ref{fig:ChampsB2} we show the shape in time of the square of magnetic field value, $B^2_f$, once filtered by the cavity low pass filter for the different sets. The birefringence effect to be measured is proportional to $B^2_f$ \cite{Agil2021}.

\subsection{Optics}

The cavity was formed by two interferential mirrors of radius of curvature of 2 meters. We have achieved a cavity finesse as high as 537~000 and an extinction ratio as low as $\sigma^2\sim 1.5\times 10^{-9}$ enabling us to have a static birefringence as low as $\Gamma\sim 5\times 10^{-5}$~rad.

For each set we have tuned the value of the static birefringence $\Gamma$ by rotating the two cavity mirrors \cite{Cadene2014}. The sign of $\Gamma$ depends on which direction we move away from the minimum corresponding to $\sigma^2$. We have fixed this sign by introducing some helium gas in the apparatus and observing its Cotton-Mouton effect which is known to be positive.

A notable feature of the new apparatus is that the vacuum tube is directly in contact with the coil and it is mechanically isolated by two bellows from the vacuum chambers.

Another novelty of these series of pulses is that we have placed at critical places of the optical set-up some absorbing materials to reduce stray light in the vacuum tanks. 

All the pulses were performed in high vacuum with pressure ranging from about $10^{-6}$ to $10^{-7}$~mbar thanks to two ionic pumps that work continuously during the magnetic pulse.

For each pulse constituting a set, we determine the finesse $F$ thanks to the photon lifetime in the cavity \cite{Cadene2014}. During each pulse, the laser delocks from the cavity, and thus the ordinary beam signal $P_t$ shows an exponential decay that we fit to deduce the time constant $\tau$ which relates to the finesse as
\begin{equation}\label{eq:finesse}
F=\frac{\pi c\tau}{L_c},
\end{equation}
where $c$ is the celerity and $L_c=2.55$~m is the length of the cavity. Therefore, for each set we have as many measure of finesse as we have of pulses.

\begin{figure}
\centering
\includegraphics[width=8cm]{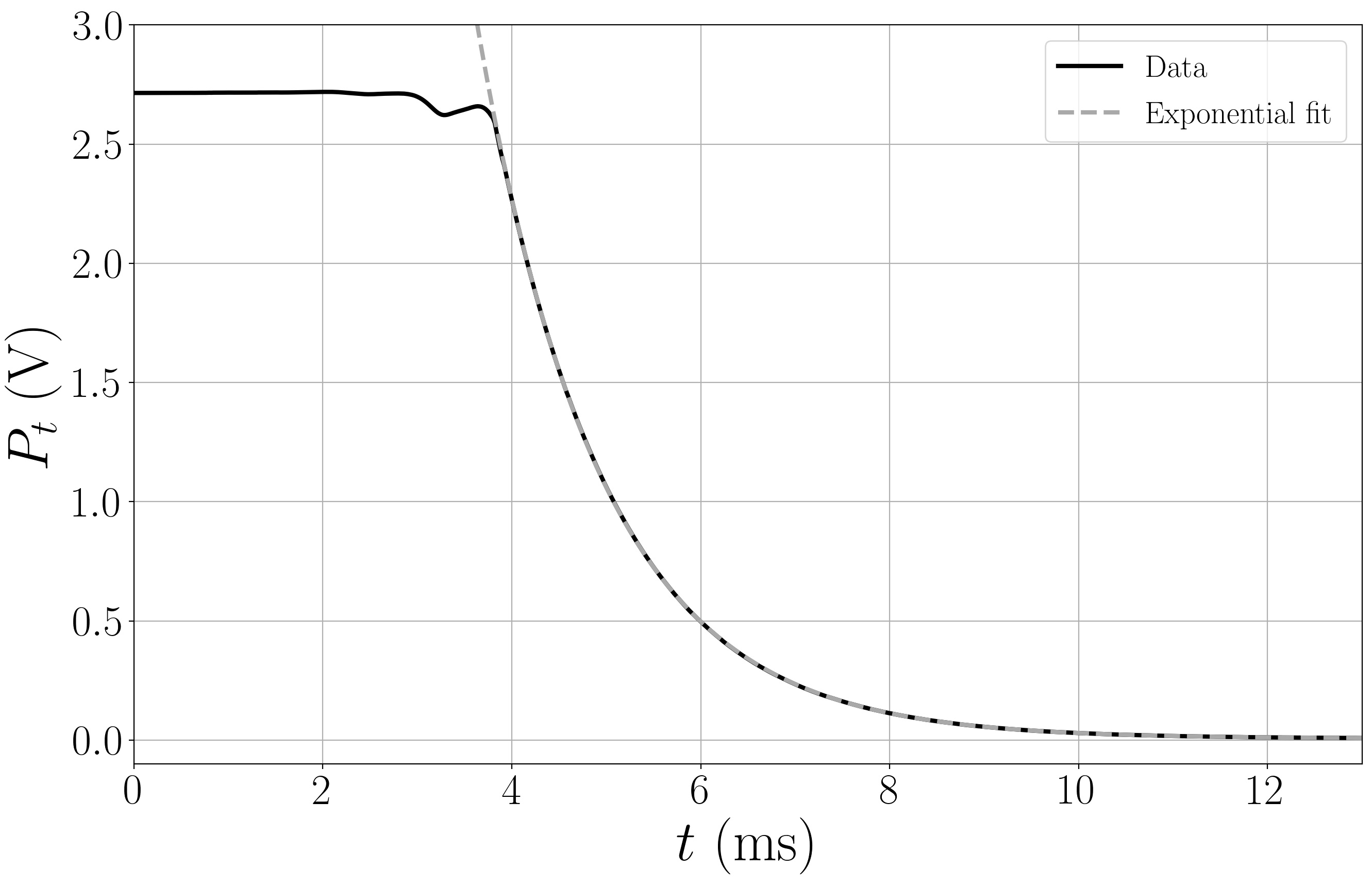}
\caption{Variation of the voltage at the ordinary beam's photodiode as a function of time, $t=0$ being the trigger of the current discharge in the coil. The pulse is extracted from the 5.5S2 set. An exponential fit is presented.}
\label{fig:Ptfitexp}
\end{figure}

In our numerical implementation, we only start the fitting procedure when the voltage of $P_t$ falls to 80\% of its mean value before the pulse. In this way, we are sure to fit only the exponential decay due to the photon lifetime and not the acoustic perturbations observed before the delocking of the laser to the cavity. An example of a fit obtained using our implementation is presented in Fig.~\ref{fig:Ptfitexp}.

\subsection{Results}

We present in the table~\ref{tab:sets}, the different data sets and their main parameters. The data sets are named according to their nominal maximum $B$ field and whether they are Long, Medium or Short pulses. Pulses with the same field but on different days have been separated in different sets because we have to align the cavity each day to reach a static birefringence of the order of $\sigma^2$, and thus, we are not in exactly the same experimental conditions.

\begin{table}
\centering
\begin{tabular}{|c|c|c|c|c|c|c|}
 \hline
 \rule{0pt}{2.5ex} Set & N$_s$ & $B^2_f$ (T$^2$) & $E_B$ (kJ) & $F$ $\times$ 10$^{-3}$ & $\sigma^2$ $\times$ 10$^{9}$ & $\nu_c$  (Hz) \\
 \hline
 \hline
 \rule{0pt}{2.5ex} 2.8L  & 51  & 6.0 & 11.0 & 537 & $1.7$ & 54.6 \\
 \hline
 \rule{0pt}{2.5ex} 3.2L1  & 69  & 8.2 & 14.3 &535 & $1.7$ & 54.9 \\
 \hline
 \rule{0pt}{2.5ex} 3.2L2  & 30  & 8.2 & 14.3 & 536 & $1.2$ & 54.8 \\
 \hline
 \rule{0pt}{2.5ex} 3.7L1  & 21  & 10.6 & 19.2 & 536 & $1.2$ & 54.8 \\
 \hline
 \rule{0pt}{2.5ex} 3.7L2  & 47  & 10.7 & 19.2 & 531 & $1.5$ & 55.3 \\
 \hline
 \hline
 \rule{0pt}{2.5ex} 2.8M  & 58  & 5.4 & 1.0 & 536 & $1.4$ & 54.8 \\
 \hline
 \rule{0pt}{2.5ex} 3.2M  & 58  & 7.3 & 14.3 & 535 & $1.4$ & 54.9 \\
 \hline
 \rule{0pt}{2.5ex} 3.7M  & 49  & 9.6 & 19.2 & 536 & $1.4$ & 54.8 \\
 \hline
 \rule{0pt}{2.5ex} 4.6M  & 45  & 14.9 & 29.6 & 537 & $1.4$ & 54.7 \\
 \hline
 \hline
 \rule{0pt}{2.5ex} 3.0S  & 69  & 5.6 & 12.6 & 443 & $3.4$ & 66.2 \\
 \hline
 \rule{0pt}{2.5ex} 3.7S  & 16  & 8.5 & 19.2 & 466 & $10$ & 63.0 \\
 \hline
 \rule{0pt}{2.5ex} 4.6S  & 16  & 13.3 & 29.6 & 466 & $16$ & 63.0 \\
 \hline
 \rule{0pt}{2.5ex} 5.5S1  & 17 & 19.0 & 42.4 & 469 & $16$ & 62.6 \\
 \hline
 \rule{0pt}{2.5ex} 5.5S2  & 32  & 18.9 & 42.4 & 476 & $4.0$ & 61.7 \\
 \hline
\end{tabular}
\caption{Parameters of the different data sets used in our analysis. $N_s$ is the number of pulses, $B_f^2$ the maximum of the filtering of the square of the magnetic field, $F$ the finesse, $\sigma^2$ the extinction ratio and $\nu_c=1/4\pi\tau$ is the cavity bandwidth.}
\label{tab:sets}
\end{table}

As far as the analysis is concerned, we have used the same procedure reported in ref. \cite{Cadene2014}. A notable feature is the separation of each pulse according to their sign of $B$ and the sign of their $\Gamma$. This separates pulses in four categories for each set. We take the mean and the standard deviation divided by the square root of the number of pulses to obtain respectively a function $Y_{\pm B,\pm\Gamma}(t)$ and the related uncertainty at each time $t$. Then we combined them to obtain the ellipticity as a function of time and finally we perform a fit by $B^2_f(t)$ to obtain $\Psi$ per T$^2$ and consequently $K_{\mathrm{CM}}$ value for the vacuum, as shown in the BMV 2014 paper \cite{Cadene2014}. This allows us to eliminate from the final ellipticity signal any systematic contribution that has not the expected symmetry with respect to the sign of B and the sign of $\Gamma$ \cite{Cadene2014}. To easily compare results to those of 2014 \cite{Cadene2014}, we first perform this analysis for each set for a time range between -3.1~ms and 3.1~ms. The only departure in our analysis from the one of 2014 is that we don't filter the ordinary beam $P_t (t)$ with the cavity bandwidth since (see \emph{e.g.} Fig.~\ref{fig:Ptfitexp}) the variation of $P_t$ during the time of analysis is of the order of 1~$\%$. This analysis simplification induces a negligible error \cite{Agil2021} to the final result. Let us also note that we did not exclude any data from our analysis.

\begin{figure}
\centering
\includegraphics[width=\linewidth]{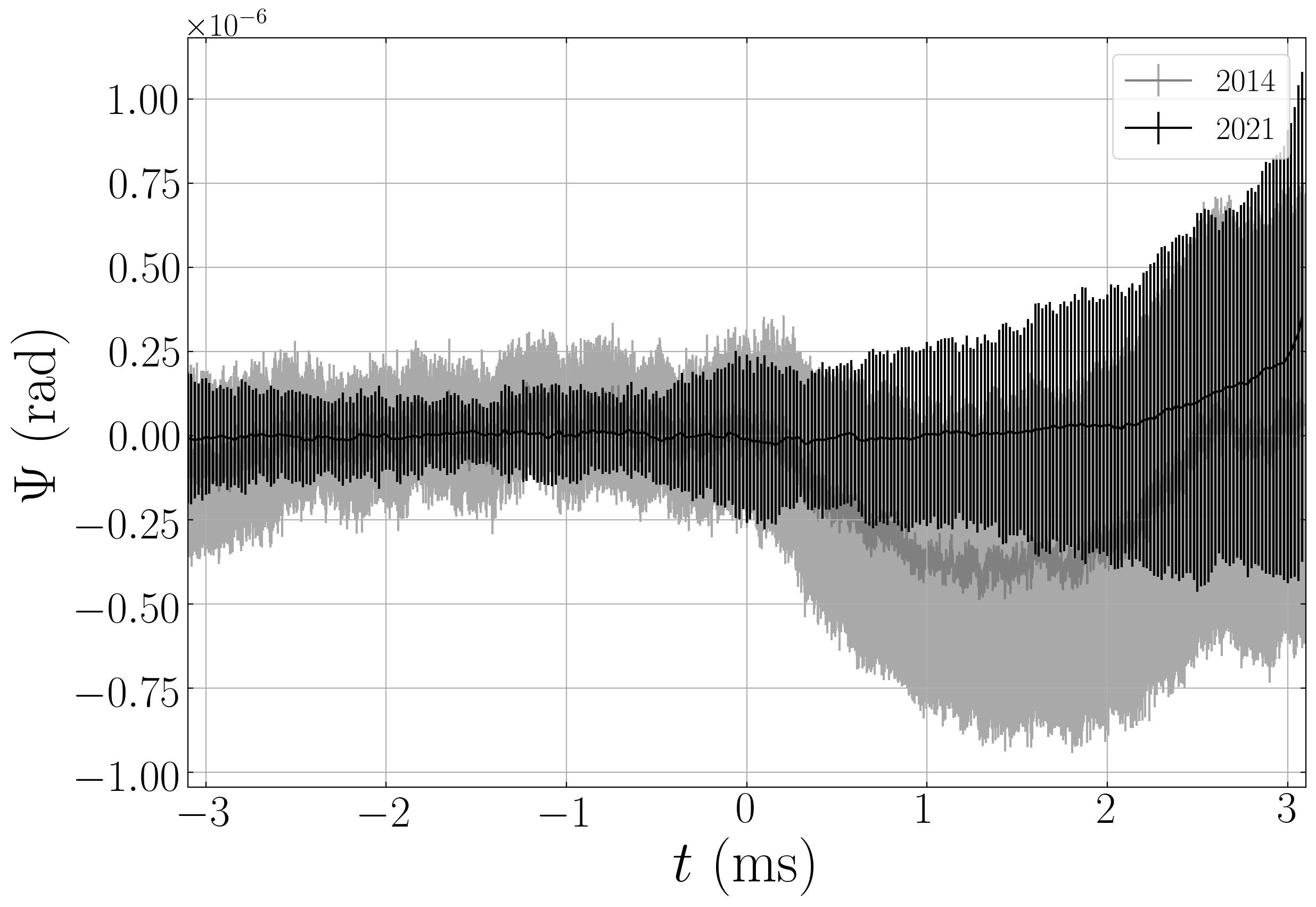}
\caption{Ellipticity in function of time of the 5.5S2 set $B^2_f L_B=10.3$~T$^2$m, and the 2014 data $B^2_f L_B=3$~T$^2$m. Data are normalized to the same number of pulses, see text. Data originally published in \cite{Cadene2014}.}
\label{fig:Comp2014}
\end{figure}

As an example, we show $\Psi(t)$ the mean of the $Y_{\pm B,\pm\Gamma}(t)$ of the 5.5S2 set in Fig.~\ref{fig:Comp2014} compared to the one of 2014. For the sake of comparison, since the 5.5S2 set consists of only 32 pulses while the 2014 results have been obtained with 101 pulses, we have reduced the error bars of the present data by a factor $\sqrt{101/32}\approx 1.8$. Both curves show the same behavior of increasing error bars but our data do not show the 2014 oscillatory systematic effect.

\begin{figure}
\centering
\includegraphics[width=\linewidth]{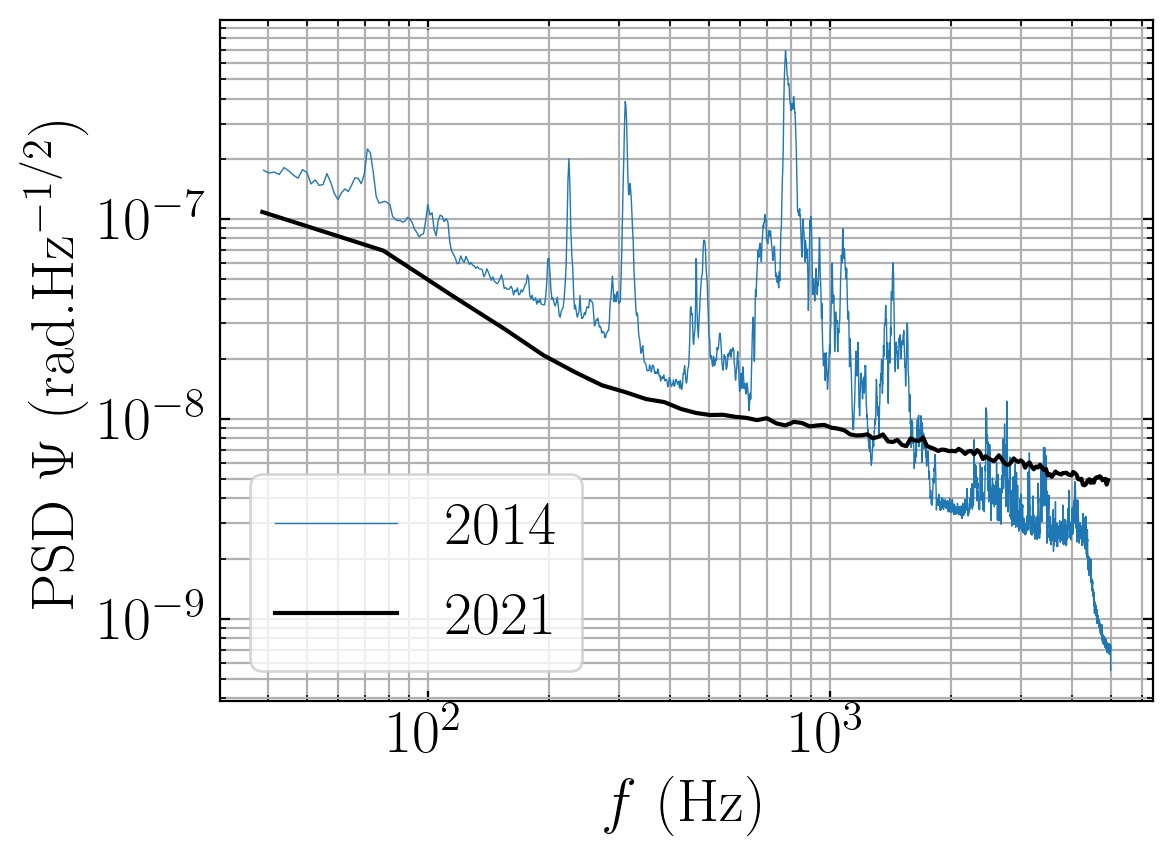}
\caption{Mean linearized Power Spectral Density before the magnetic pulse of every pulse. For reference we show the PSD origininally published in \cite{Cadene2014}.}
\label{fig:PSD}
\end{figure}

Moreover, as we show in Fig.~\ref{fig:PSD}, the mean linearized Power Spectral Density before the magnetic pulse confirms that mechanical resonances have disappeared with respect to the Power Spectral Density reported in 2014 \cite{Cadene2014}.

To study the impact of a pulse on noise, we fit the ellipticity also before the magnetic pulse from $t=-3.1$ to $t=0$~ms where any noise coming from the pulse itself is obviously absent. This value gives information on the result that we would have obtained if the experiment were limited by the optical noise.

Then, we compare the ellipticity uncertainty before the magnetic pulse $\Delta\Psi_{opt}$ and the one obtained during the pulse $\Delta\Psi_{tot}$.
Before the magnetic pulse, the uncertainties arise only from the optical noise of the experiment. During the pulse however, we have the added noise of the pulse, $\Delta\Psi_{B}$, that contributes to the total uncertainty $\Delta\Psi_{tot}$.  We can estimate $\Delta\Psi_{B}$ by
\begin{equation}
\Delta\Psi_{B}=\sqrt{\Delta\Psi_{tot}^2-\Delta\Psi_{opt}^2}.
\end{equation}

We compute the ellipticity uncertainty for each set, we multiply it by the square root of the number of pulses $N_s$ of the set to obtain the noise $\Delta\Psi_{B}^{n}$ associated to a single pulse. The final $\Delta\Psi_{B}$ is obtained by multiplying the single pulse noise to the corresponding $B^2_{f,m}$ which is the $B_f^2$ reached at the end time of analysis $t_{an}$.

\begin{figure}
\centering
\includegraphics[width=\linewidth]{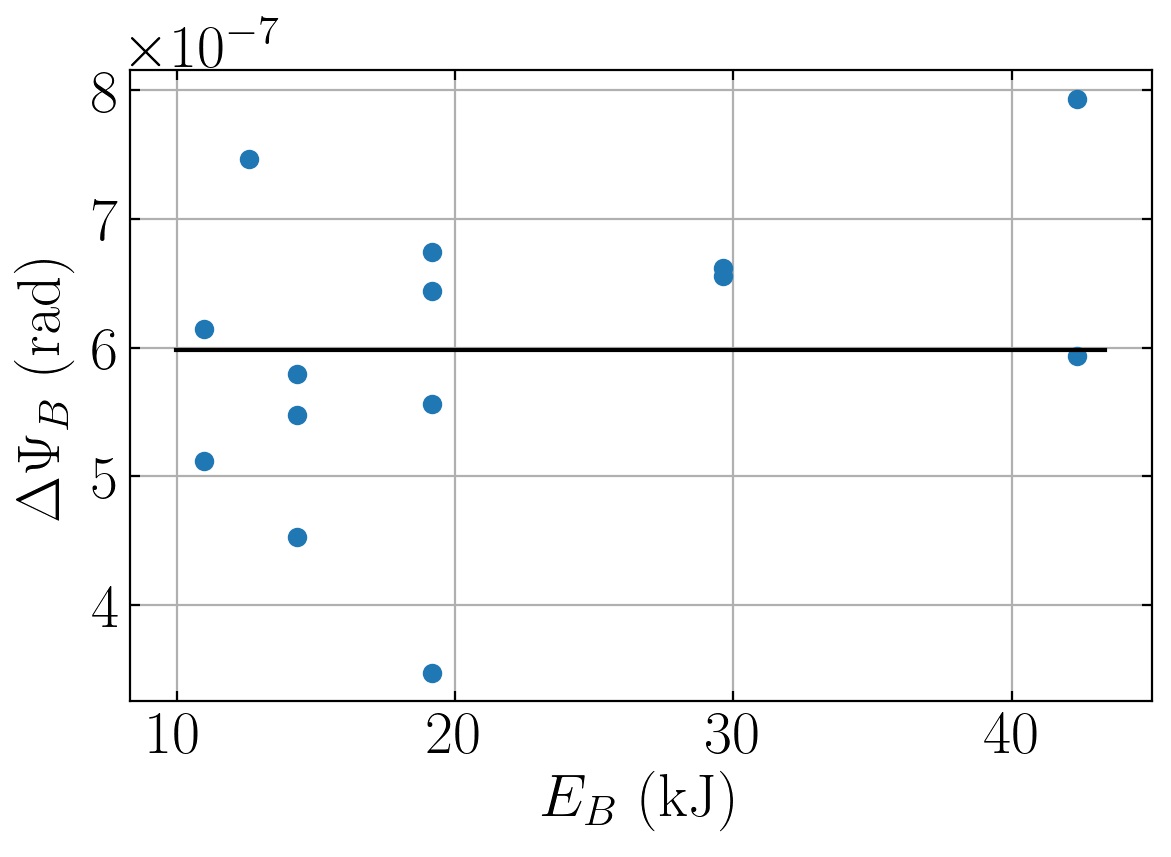}
\caption{Noise of the pulse $\Delta\Psi_{B}$ as a function of the magnetic energy delivered by the coil. The full line showing the mean of the points, $\overline{\Delta\Psi}_{B}=6.0\times 10^{-7}$~rad.}
\label{fig:DPsivsB2}
\end{figure}

In Fig.~\ref{fig:DPsivsB2} we show $\Delta\Psi_{B}$ as a function of the energy injected into the coil. Somewhat surprisingly, we see that $\Delta\Psi_{B}$ is constant in our range of energies, $\overline{\Delta\Psi}_B= 6\times 10^{-7}$~rad, indicating that we gain by increasing the magnetic field value since we increase the signal to noise ratio. It is a clear indication that we need to pulse the magnet to its maximum energy, that is a magnetic field of 11~T. 

\begin{figure}
\centering
\includegraphics[width=\linewidth]{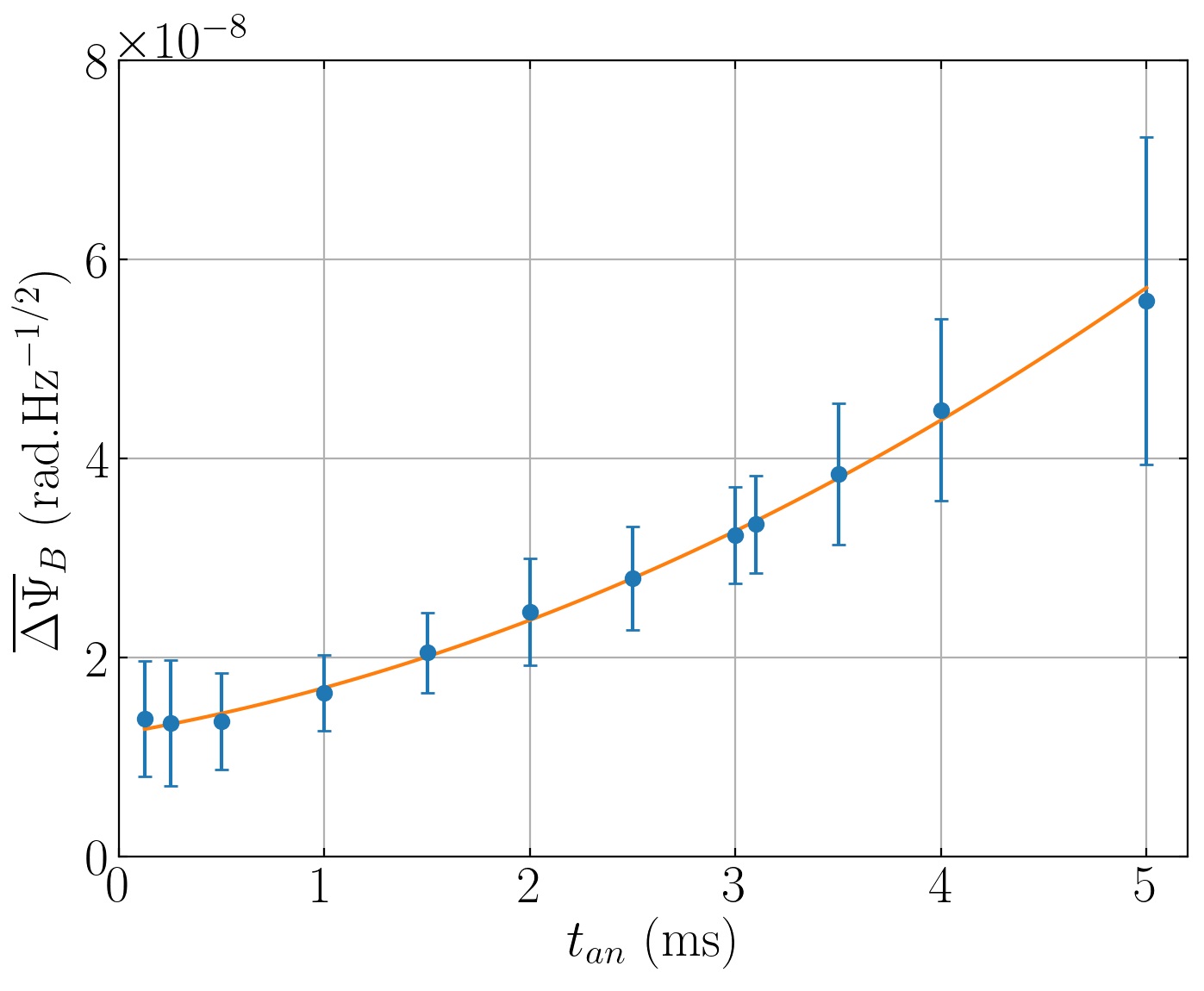}
\caption{Mean of the magnetic noise $\overline{\Delta\Psi}_B$ as a function of the analysis time $t_{an}$. A parabolic fit is also shown.}
\label{fig:Dpsivsta}
\end{figure}

To study the dependence of $\overline{\Delta\Psi}_{B}$ on $t_{an}$ we calculate it for different analysis time from 0.125~ms to 5~ms in Fig.~\ref{fig:Dpsivsta} where we show the results obtained in rad/$\sqrt{\mathrm{Hz}}$ together with a parabolic fit.

This results indicate that soon after the beginning of the pulse ($\approx 1$~ms), whatever is the energy injected in the coil, the noise induced by the coil itself increases almost linearly, confirming our Monte Carlo results \cite{Agil2021}. Finally, we calculate the $K_{CM}$ constant corresponding to the two 5.5~T sets. A reasonably well accurate and precise value was obtained with a $t_{an}=2.7$~ms, $K_{CM}=(0.2\pm 1.0)\times 10^{-20}$~T$^{-2}$. The analysis between $-t_{an}$ and 0 gives $K_{CM}=(0.1\pm 3.7)\times 10^{-21}$~T$^{-2}$. The total number of pulses of the two sets is 49, see table~\ref{tab:sets}.

\begin{figure}
\centering
\includegraphics[width=\linewidth]{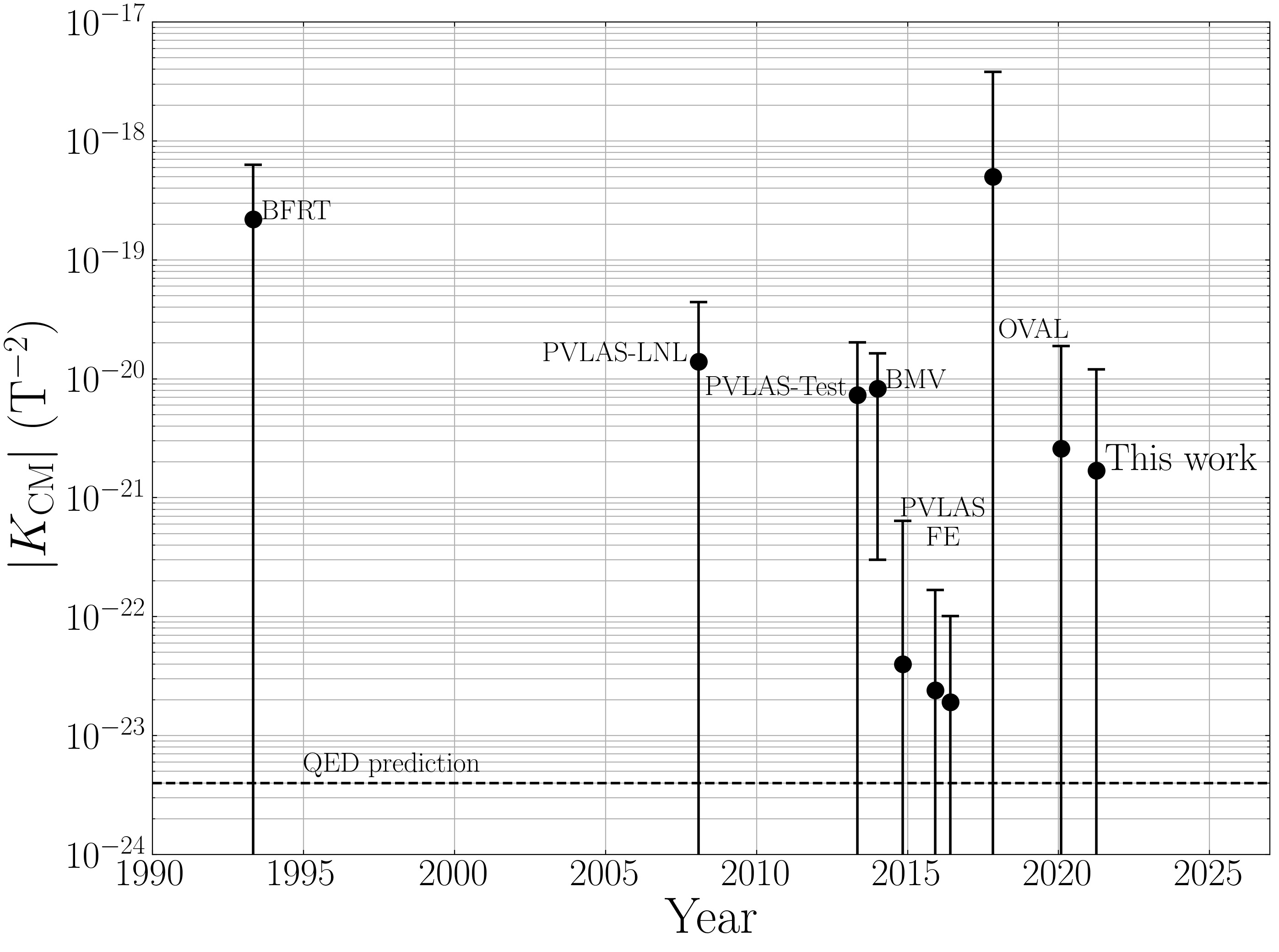}
\caption{Measurements of vacuum magnetic birefringence across the years, errors bars are represented with a coverage factor $k=3$. Absolute values are derived from data originally published in the following references BFRT \cite{Cameron1993}, PVLAS-LNL \cite{PVLAS2008}, PVLAS-Test \cite{PVLAS2013}, BMV \cite{Cadene2014}, PVLAS-FE \cite{PVLAS2014}\cite{PVLAS2016}\cite{PVLAS2020} and OVAL \cite{OVAL}\cite{OVALThesis}.}
\label{fig:AllRes}
\end{figure}

\section{Conclusions}
As for the coil itself, some potential improvement in the design can be made to increase the efficiency or the ergonomics. First an extra cooling to remove the heat from the box will be installed. Actual average cooling power due to natural convection and conduction is lower than 50~W and can easily be increased with a forced flow  of air or cold nitrogen gas inside the box. A second modification is to optimize the pulse duration, probably by shortening it, either by adapting the actual capacitor bank or using another one available at LNCMI, without lowering the maximum field. A next step could be a modification in the design itself. For example, it is possible to cut the copper foil to maintain the same current density all along the winding. It will concentrate the current density closer to the laser beam path increasing the efficiency, \textit{\emph{i.e.}.} $B/I$ by a factor around 1.3 without increasing the risk of buckling around the optical access. 

As for its use in the BMV experiment, the Foil Coil proved to be a real improvement and more generally for magnetic linear birefringence measurements. The absence of a cryostat simplifies the whole apparatus. Last but not least, it also reduces the costs of fabrication and operation. On the other hand, nothing prevents us from improving the cooling of the Foil Coil to gain a factor of two with respect to the 12.5~T field, however, it is not clear that the gain of a factor 4 in the effect is worth the big technical effort for cryogenic cooling. 

As far as preliminary physics results, our best value has been obtained with the short pulses at 5.5~T and an analysis time of 2.7~ms where we reached a maximum field similar to 2014 and obtained $K_{\mathrm{CM}}=(0.2 \pm 1.0)\times 10^{-20}$~T$^{-2}$, which is about the same as the value obtained in  2014, if we consider that this has been obtained with half the number of pulses, but much more accurate. This value is shown with other reported results in Fig.~\ref{fig:AllRes}. Indeed, unlike in the 2014 results, we do not observe a systematic effect. The insertion of the new magnet in the apparatus has been very successful. On the other hand,  we did not gain yet in precision as much as we should have, despite the fact that we have a $B^2_f L_B$ three time as high and a mean PSD about two times lower. $\Gamma$ is also lower than in 2014, and this should also reduce the optical noise. This clearly indicates that optics optimization has to be pushed further. This is why we have not pushed the field strength higher than the 2014 one, our reference point. We considered that it was not worthwhile to prematurely age the coil at this stage.

Nevertheless, we are still limited by the noise induced by the pulse itself. A signature of this noise is in the increasing error bars during the pulse. Actually, $K_{\mathrm{CM}}^{opt}$ is 3 times better than $K_{\mathrm{CM}}$ confirming that the noise induced by the pulse itself is still limiting us. 

The magnet commissioning and the first results look encouraging. We are currently working to diminish the overall noise by better acoustically insulating the apparatus from the coil. On a long term perspective we are also studying how to suspend in vacuum the optical tables holding mirrors and polarizers. The new generation of the BMV experiment has just begun, and there is very much room for improvement.

\section*{Acknowledgments}
We thank all the technical staff of the LNCMI. We thank all the members of the BMV collaboration. We specially thank M. Hartman and all the authors of \cite{Cadene2014}. We also thank D. Maude for the careful proofreading of this manuscript.

\section*{Data Availability Statement}

The data that support the findings of this study are available from the corresponding author upon reasonable request.




\end{document}